\documentstyle[revtex]{aps}
\math-with-secnums
\begin{document}
\def\partder#1#2{{\partial #1\over\partial #2}}
\begin{title} EXACT 1-FERMION LOOP CONTRIBUTIONS
		IN 1+1 DIMENSIONAL SOLITONS
\end{title}
\author{J. R. Shepard, C. E. Price and T. C. Ferr\'ee}
\begin{instit} Department of Physics, University of Colorado, Boulder, CO 80309
\end{instit}
\begin{abstract}
We find solutions to the 1+1 dimensional scalar-only linear sigma
model.  A new method is used to compute 1-fermion loop
contributions exactly and agreement with published results employing
other methods is excellent.
A renormalization scheme which differs from that commonly used in such
calculations but similar to that required in 1+3 dimensions is also
presented.  We compare
``kink'' {\it versus} ``shallow bag'' solutions paying careful attention
to the implications of the 1-fermion loop contributions for the stability of
the former.  We find that, for small fermion multiplicities,
self-consistent shallow bag solutions are always more bound than their
metastable kink counterparts.  However, as the fermion multiplicity
increases, shallow bags evolve into kinks which eventually are the
only self-consistent configurations.  This situation is qualitatively
the same for the two renormalization schemes considered.  When we construct
``baryons'', each containing three fermions, the kink configuration is
typically more bound than the shallow bag when 1-fermion loop
contributions are included.
\end{abstract}

\newpage
\narrowtext
\section{INTRODUCTION}

	In the absence of a sufficiently tractable QCD-based description
of low-energy hadronic phenomena, there has been considerable interest
in developing relatively simple field theoretic models which possess
some of the important features one might anticipate in the full QCD
treatment.  For example, the linear sigma model\cite{js,gl,lee,candl,col}
has been studied for three decades because of its
conceptual simplicity, dynamical richness and ``built-in'' chiral
symmetry.
In this model, fermions are massless at the level of the Lagrangian
and acquire mass through spontaneous symmetry breaking in the
vacuum.  One of the more intriguing features of the model is the
existence of solutions -- at least in the classical limit for the scalar
field -- for which the fermions are strongly confined to a finite region
where the scalar field vanishes.  The fermions are then massless in this
region where the broken chiral symmetry is approximately restored.
The confinement anticipated in the full QCD solutions is also approximated
in that the energy
required to remove a fermion from the finite region where it is bound
roughly equals the spontaneously generated fermion mass and is
thus large on the low-energy scale the model seeks to describe.  These
solutions -- often referred to as ``kink solutions'' --
have been examined in detail by, {\it e.g.}, the SLAC Bag
Collaboration\cite{sbag} in an effort to describe hadronic structure
and by a variety of others (see, {\it e.g.}, Refs.~\cite{lw,bog} )
to study``abnormal phases'' of nuclear matter.

	Almost from the very beginning of these studies, there has been
speculation about the persistence of such solutions when
``one-loop'' vacuum
contributions are included.\cite{candl,lw}  Indeed, there is
current interest\cite{ahh,dlst,bandn} in this question.  Such
``semi-classical''
analyses are complicated by the fact that the kink or deep-bag solutions
are of course non-uniform and therefore the treatment of 1-loop vacuum
contributions is considerably harder than for uniform systems.
Furthermore, the kink is in some sense {\it maximally} non-uniform since
the spontaneously generated fermion mass {\it changes sign} over a
spatial interval comparable to the Compton wavelength associated
with the maximum fermion mass.  Hence calculations which employ, {\it e.g.},
derivative expansions of an effective action\cite{bandn} must be
scrutinized very carefully -- especially since, as will be seen below,
energy differences between kink and more conventional
solutions can be quite small. More inspiring of confidence are
approaches which treat 1-loop vacuum effects exactly.  For example, Campbell
and Liao\cite{candl} presented early calculations of 1-fermion loop
contributions to the kink energy in 1+1 dimensions.  However, their
treatment was restricted to a single special set of
coupling constants.  Li, Perry and Wilets\cite{lpw} and, more
recently, 
Wasson\cite{wass} have
reported similar exact calculations.  Wasson has furthermore presented
results for a variety of couplings and for which the 1-fermion loop
contributions are treated in a fully self-consistent manner.  While some
of these authors have focussed exclusively on calculational aspects of
the problem, others\cite{candl,bandn} have concluded that 1-fermion loop
contributions render the kink solution either absolutely unstable or
metastable with respect to other ({\it e.g.}, shallow bag) solutions.  In
the present work, we pursue these questions further using a
newly-developed method for exact treatment of 1-loop
effects.\cite{fer}  Like Wasson, we find self-consistent solutions
for a variety of couplings.
In fact, with our method, we are able to reproduce his numerical results
exactly (as well as those of Campbell and Liao) and with great
computational efficiency.  We also go beyond these studies by
including counter terms which, though finite in one spatial dimension,
are required to eliminate divergences in three dimensions.
Our objectives here include (1) a detailed presentation of our
calculational method and (2) a study of the relative stability of
semi-classical kink and shallow bag solutions for various sets of input
parameters.

\section{REVIEW OF THE THEORETICAL MODEL}

	We assume the following Lagrangian in 1+1 dimensions:
\FL
\begin{equation}
  {\cal L}=\bar\psi i\gamma\cdot\partial\psi + g\phi\bar\psi\psi
	 + {1\over 2}\partial\phi\cdot\partial\phi - U(\phi)
	+ {\cal L}_{ctc}  \label{ba}
\end{equation}
where $\psi$ is the fermion field, $\phi$ is the scalar field and where
${\cal L}_{ctc}$ is the counter-term Lagrangian whose exact form will be
specified below.  The scalar potential is
\begin{equation} U(\phi)={\lambda\over 4}(\phi^2-f^2)^2. \label{bb}
\end{equation}
This is of course the Lagrangian of the linear sigma model without the
pseudoscalar field and it thus possesses discrete rather than continuous
chiral symmetry.  For a uniform system, in the absence of coupling to
the fermion field ($g=0$), the classical scalar field satisfies
\begin{equation} \phi_c^2=f^2. \label{bc} \end{equation}
There also exists a higher energy non-uniform ``kink'' solution
\begin{equation} \phi_c(x)=f\ {\rm tanh}\biggl(\sqrt{{\lambda\over 2}} f\ x
	\biggr) \label{bxd} \end{equation}
which ``interpolates'' between the two uniform solutions, namely
$\phi_c=\pm f$.  In general, the classical field dynamically generates
both fermion and scalar masses.  If we write $\phi=\sigma + f$, then
\begin{eqnarray} {\cal L}\rightarrow \bar\psi \bigl[i\gamma\cdot\partial
	&-& (m+S) \bigr]\psi
	+{1\over 2} \partial\sigma\cdot\partial\sigma \nonumber \\
	&-& {1\over 2}
	\mu^2_0\sigma^2 - V(\sigma) + {\cal L}_{ctc} \label{be} \end{eqnarray}
where
\begin{eqnarray} m=-gf{\rm ,}\quad &S&=-g\sigma{\rm ,}\quad
	\mu^2_0=2\lambda f^2{\rm ,} \nonumber \\
	\quad && V(\sigma)=\lambda\bigl({{\sigma^2}\over{4}}+\sigma f
	\bigr)\sigma^2.
	\label{bff} \end{eqnarray}
The Euler-Lagrange equations are
\begin{equation} (i\gamma\cdot\partial -m^*)\psi=0 \label{bg} \end{equation}
and
\def\sqr#1#2{{\vcenter{\vbox{\hrule height.#2pt \hbox{\vrule width.#2pt
height #1pt \kern#1pt \vrule width.#2pt} \hrule  height.#2pt}}}}
\def\abox{\mathchoice\sqr{10}5\sqr{10}5\sqr{7}3\sqr{4}3}
\begin{equation} (\abox+\mu^2_0)\sigma + V'(\sigma) = g\bar\psi\psi \label{bh}
\end{equation}
where $m^*=m+S=-g\phi$ and $V'(\sigma)\equiv{{dV}\over{d\sigma}}$.
Eqn.~\ref{bh}\ is solved subject to the approximation that
$\bar\psi\psi$ is replaced by its expectation value
\begin{equation}
\bar\psi\psi\rightarrow<\bar\psi\psi>=\rho_s=\sum_{occ'd}\bar\psi_i
	\psi_i  \label{bi} \end{equation}
and that fluctuations of the shifted scalar field are ignored in which
case $\sigma$ becomes a $c$-number.
Since the sum in Eqn.~\ref{bi}\ runs over all occupied fermion states including
the negative energy sea, $\rho_s$ is divergent and must be renormalized.
Then Eqns.~\ref{bg}\ and \ref{bh}\ represent a set of coupled
equations to be solved self-consistently.  Deferring for now a
discussion of the renormalization of $\rho_s$, we outline our method of
solution of the coupled equations of motion.  Details appear in
Ref.~\cite{fer}.

	Our solutions for Eqs.~\ref{bg}\ and \ref{bh}\ assume antiperiodic
boundary conditions for the fermion spinors which then imply the
densities and scalar fields are periodic.  The free spinors ($m^*=m$)
are
\FL
\begin{equation} \varphi^{e+}_j=N_j \left(\matrix{\cos k_j x \cr
				{{i k_j\sin k_j x}\over{E_j^0+m}} \cr}
				\right) {\rm ,}\qquad
   \varphi^{o+}_j=N_j \left(\matrix{\sin k_j x \cr
				{{-i k_j\cos k_j x}\over{E_j^0+m}} \cr}
				\right) \end{equation}
\FL
\begin{equation} \varphi^{e-}_j=N_j \left(\matrix{{-{k_j\cos k_j
x}\over{E_j^0+m}} \cr
				    i\sin k_j x \cr}
				\right) {\rm ,}\qquad
   \varphi^{o-}_j=N_j \left(\matrix{{{k_j\sin k_j x}\over{E_j^0+m}} \cr
				    i\cos k_j x \cr}
				\right) \label{bj} \end{equation}
where $e (o)$ refers to even (odd) parity {\it of the upper components}
and where $+ (-)$ implies positive (negative) energy solutions and
$k_j\equiv(2j-1)\pi/a$, $a=$ box length and $E_j^0\equiv
+\sqrt{k^2_j+m^2}$.  The normalization factor is given by $N_j^2=
[(E_j^0+m)/2E_j]\cdot(2/a)$.  Stationary solutions to the Dirac equation
(Eqn.~\ref{bg}) are obtained separately for even and odd states  by assuming
\begin{equation} \psi_{n'}=\sum^N_{j=1}\bigl[ a^+_j(n')\varphi^+_j + a^-_j(n')
	\varphi^-_j \bigr] \label{bk} \end{equation}
where $N$ specifies the size of the truncated free basis and where
the parity label has been suppressed.

	The scalar field will have the same form as the density which is
bilinear in the spinors.  Hence,
\begin{equation} \phi(x)=\sum_{n=0}^{2N-1} \phi_n \cos {{2n\pi x}\over{a}}
\end{equation}
or, equivalently,
\begin{equation} \sigma(x)=\sum_{n=0}^{2N-1} \sigma_n \cos {{2n\pi x}\over{a}}
	\label{bl} \end{equation}
where the upper limit is a consequence of the truncated
basis.  From the definition of the scalar potential given above, it
is clear that we also have
\begin{equation} S(x)=\sum_{n=0}^{2N-1} S_n \cos {{2n\pi x}\over{a}} \label{bm}
\end{equation}
where $S_n = -g\sigma_n$.  Matrix elements of $S$ for the free basis are
simply evaluated analytically and the solutions to the Dirac equation
are readily obtained by solving a $2N\times 2N$ matrix eigenvalue
problem for each parity.  This yields energy eigenvalues, $E^\pi_{n'}$, and
eigenvectors, {\it i.e.}, the $a^{\pm,\pi}_j(n')$.  These solutions may
then be used to construct the fermion scalar density:
\begin{equation} \rho_s(x)=\sum_{occ'd} \bar\psi_{n'}\psi_{n'}\rightarrow
	\sum_{n=0}^{2N-1} \rho_n \cos {{2n\pi x}\over{a}}. \label{bn} \end{equation}
We again observe that $\rho_s$ must be renormalized; the method used
here will be discussed in detail below.  Solution of the equation of
motion for the scalar field, Eqn.~\ref{bh} , in the static limit is now very
simple.  For each Fourier coefficient (Eqn.~\ref{bl} ) we find
\begin{equation} \sigma_n=-(g\rho_n + V'_n)/\biggl[\biggl(
	{{2n\pi}\over{a}}\biggr)^2
	+\mu^2_0\biggr] \label{bo} \end{equation}
where the coefficients $V'_n$ are defined via
\begin{equation} V'[\sigma (x)]=\sum_{n=0}^{2N-1} V'_n \cos {{2n\pi x}\over{a}}
\end{equation}
and may readily be constructed from the coefficients $\sigma_n$.  The
fermion and scalar equations of motion are solved iteratively until
self-consistency is achieved.  As will be discussed below, this method
of solution is quite efficient in that only modest basis sizes are
required and convergence to self-consistency is rapid.  Further
efficiency is realized in that no complicated algorithms are required to
ensure that individual states are not ``lost'' in solving for the
fermion spectrum.  The matrix diagonalization method automatically
provides the entire (truncated) spectrum even for the unusual potentials
encountered in finding kink solutions.  In contrast, using standard
$x$-space methods such as Runge-Kutta, great care in setting initial
conditions at $x=0$ and $a/2$ and in ``counting levels'' -- which are
found one at a time -- is required to accomplish the same result.
Beyond this, the matrix methods ensure that an important ``local
completeness'' relation\cite{fer} -- difficult and perhaps even
impossible to incorporate in other methods -- is automatically
satisfied.

\section{RENORMALIZATION METHODS}

	Our method of renormalization begins with the evaluation of the
{\it unrenormalized} energy density of the sea:
\begin{equation}{\cal E}_V (x)\equiv \sum_{n'=1}^N \sum_\pi\ E^{(\pi,-)}_{n'}\
	\bar\psi^{(\pi,-)}_{n'}(x)\ \gamma^0\
	\psi^{(\pi,-)}_{n'}(x) \label{xa} \end{equation}
where $E^{(\pi,-)}_{n'}$ and $\psi^{(\pi,-)}_{n'}$ are the negative
energy eigenvalues and eigenspinors obtained via the methods of the
previous section.  This density is large but finite due to the
finiteness of our discrete basis.  In the discussion to follow, we
expand ${\cal E}_V$ in powers of $\sigma (x)$ (or, equivalently, in
powers of $S(x)$) -- not to evaluate ${\cal E}_V (x)$ since that
quantity is already in hand -- but to identify the counter term
contributions, ${\cal E}_{ctc}(x)$,
required to render ${\cal E}_V (x)$ finite in the limit
of very large basis size.  This procedure then yields ${\cal E}_{loop}
={\cal E}_V - {\cal E}_{ctc}$, the 1-fermion loop contribution to
the total energy density.  Expressions for other quantities such as the
1-loop contribution to the fermion scalar density, $\rho_s$,
follow immediately.

	To start we observe that,
with the fermion basis truncated at momentum $\Lambda$, the
unrenormalized energy density due to the sea is, {\it for a uniform
system with} $\phi^2=f^2$ (or, equivalently, for $S=0$),
\begin{equation} {\cal E}^{(0)}_V (x)=-\int^{+\Lambda}_{-\Lambda}\
{{dk}\over{2\pi}}\
	E^0_k \label{ca} \end{equation}
where $ E^0_k\equiv +\sqrt{k^2+(gf)^2}=\sqrt{k^2+m^2}$. Upon
discretizing,
\begin{equation} {\cal E}^{(0)}_V\rightarrow{1\over a} \sum_{i=-N+1}^N (-E^0_i)
	\label{cb} \end{equation}
where $E^0_i\equiv\sqrt{k^2_i + (gf)^2}$ and $k_i\equiv(2i-1)\pi/a$.
Note that we here refer to a free basis with positive {\it and} negative
wavenumbers rather than the basis of Eqn.~\ref{bj}\ characterized
by states with positive wavenumbers only but with well-defined parity.
For quantities involving summations over {\it all} negative energy
states, the two bases may be used interchangeably.

	The {\it shift} in this energy density due to $S(x)\neq 0$ is,
schematically,
\begin{equation} {\cal E}_V (x)-{\cal E}^{(0)}_V (x) 
	=\sum_{n=1}^\infty {1\over n}{\rm Tr} \bigl[SG_0 \bigr]^n
	\label{cc} \end{equation}
where the ``free'' fermion propagator is
\begin{equation} G_0(x-y;\omega)=\int_{-\infty}^{+\infty} {{dp}\over{2\pi}}\
	e^{-ip(x-y)}\ G_0(\omega ,p) \label{cd} \end{equation}
with
\begin{equation} G_0(\omega ,p)=G_0(p^\mu)={{\theta(\Lambda-|p|)}\over
	{\gamma^\mu p_\mu - m + i\epsilon}}. \label{ce} \end{equation}
To proceed, we {\it temporarily} assume the following form for the
counter-term Lagrangian:
\begin{equation} -{\cal L}_{ctc}={1\over 2} \xi\ \partial\phi\cdot\partial\phi\
+\
	\sum_{n=1}^4 {{\alpha_n}\over{n!}}\phi^n \label{cf} \end{equation}
or, equivalently, since a vacuum expectation value (VEV) subtraction is
always understood,
\begin{equation} -{\cal L}_{ctc}={1\over 2} \xi\ \partial S\cdot\partial S +
\sum_{n=1}^4
	{{\alpha_n}\over {n!}}\ S^n \label{cg} \end{equation}
where, of course, we have redefined the counter term coefficients.
The key to our method is that we
fix these coefficients so as to cancel to the greatest possible degree
the leading order terms in the expansion of the shift in the energy density
given in Eqn.~\ref{cc}.  This identifies our renormalization point as the
uniform system with $S=0$.

	We now consider the first several terms in Eqn.~\ref{cc}\ beginning
with
\widetext
\begin{eqnarray} {\cal E}^{(1)}_V (x)\equiv {\rm Tr}\ SG_0 && \rightarrow
	  \int_{-\infty}^{+\infty}{{d\omega}\over{2\pi i}}\quad
	  e^{+i\omega 0^+}\ {\rm Tr}\ \bigl[S(x)\ G_0(x-x;\omega)
	  \bigr] \nonumber \\
	&&=S(x)\ \int_{-\Lambda}^{+\Lambda} {{dp}\over{2\pi}}\
	  \biggl( -\partder{E^0_p}{m} \biggr) \label{ch} \end{eqnarray}
\narrowtext
After  discretization
\begin{equation} {\cal E}^{(1)}_V (x) \rightarrow
	S(x)\cdot {1\over a} \sum_{i=-N+1}^N \biggl(
	  -\partder{E_i^0}{m} \biggr) \label{ci} \end{equation}
and we conclude that, consistent with our requirement of maximal
cancellation, the  $\alpha_1$ coefficient in ${\cal L}_{ctc}$
(Eqn.~\ref{cg} ) is
\begin{equation} \alpha_1 = {1\over a} \sum_{i=-N+1}^N \biggl(
-\partder{E^0_i}{m}
	\biggr)=\rho^{(0)}_V \label{cj} \end{equation}
which we also identify as the lowest order unrenormalized scalar density
contribution from the sea.  
It is crucial to note that we have
implicitly assumed that $\alpha_1$ is independent of $x$ as a counter
term {\it coefficient} must be.  However, in its original form based on
Eqn.~\ref{ch} , this quantity is
\begin{equation} \alpha_1=\sum_{\pi,j}\ \bar\varphi^{\pi,-}_j (x)
	\ \varphi^{\pi,-}_j (x) \end{equation}
and the $x$-dependence explicit in the sum only vanishes because of the
``local completeness'' of the basis.\cite{fer}  Referring to the
expressions for the free basis states appearing in Eqn.~\ref{bj} , we see
that, at its most fundamental level, ``local completeness'' emerges in the
present case by virtue of the identity $\cos^2 k_j x + \sin^2 k_j x =1$,
independent of $x$ (or, equivalently, because $|exp(ik_j x)|=1$ for all
$x$ since the $k_j$ are real).  In
solving the Dirac equation by diagonalizing within our truncated basis,
we ensure that our fully interacting basis, {\it i.e.}, the $\psi_{n'}$
of Eqn.~\ref{bk} , satisfies similar local completeness relations.  The fact
that such relations are satisfied in practice with very great numerical
precision is the key to our ability to evaluate renormalized quantities
reliably by straightforward subtraction of counter terms.

	The next term is
\widetext
\begin{eqnarray} {\cal E}^{(2)}_V (x)&=&{1\over 2} S(x)
	\int_{-\infty}^{+\infty}
	{{d\omega}\over{2\pi i}}\quad
	  e^{+i\omega 0^+} \int_{-\infty}^{+\infty} dy\quad S(y)\
	  {\rm Tr}\ \bigl[ G_0 (x-y;\omega)\ G_0 (y-x;\omega) \bigr]
	  \nonumber \\
	&=&{1\over 2} S(x)\ \rho^{(1)}_V (x) \label{ck} \end{eqnarray}
which defines the next order sea contribution to the scalar density.
After some algebra we find
\begin{eqnarray} \rho^{(1)}_V (x) =\int_{-\infty}^{+\infty}
	  {{dp}\over{2\pi}} \int_{-\infty}^{+\infty} {{dp'}\over{2\pi}}
	&&\int_{-\infty}^{+\infty} dy\quad S(y)\ e^{-i(p-p')(x-y)}
	\nonumber \\
	&& \times\theta(\Lambda-|p|)\ \theta(\Lambda-|p'|)\
	  {\cal F}(p,p') \label{cl} \end{eqnarray}
where, dropping the superscripts on, {\it e.g.}, $E^0_p$,
\begin{equation}
	{\cal F}(p,p')\equiv-{{E_p E_{p'} + pp' -m^2}\over{(E_p + E_{p'})
	E_p E_{p'}}}. \label{cm} \end{equation}
We now define $q\equiv p-p'$ and $k\equiv(p+p')/2$ and expand:
\begin{equation}
    {\cal F}(p,p')={\cal F}[(k+q/2),(k-q/2)]={\cal F}_0 (k) + {\cal F}_2 (k)
	q^2 + {\cal O} (q^4) \label{cn} \end{equation}
where
\begin{equation} {\cal F}_0 (k) 
	=-\partder{^2 E_k}{m^2} \end{equation}
and
\begin{equation} {\cal F}_2 (k)= {{5k^2m^2}\over{8 E^7_k}}. \end{equation}
Now
$ \int_{-\infty}^{+\infty} dy\ S(y)\ e^{iq(x-y)}\rightarrow\tilde S(q)
	\ \cos qx $ since $S(y)$ is a real, even function.  Then
\begin{eqnarray} \rho^{(1)}_V (x) =\int_{-\infty}^{+\infty} {{dk}\over{2\pi}}
	\int_{-\infty}^{+\infty} {{dq}\over{2\pi}}\ \tilde S(q)\ &&\cos qx
	\ \theta(\Lambda-|k+q/2|)\ \theta(\Lambda-|k-q/2|) \nonumber \\
	&&\times\bigl[ {\cal F}_0 (k) + {\cal F}_2 (k) q^2 \bigr]
	\label{co} \end{eqnarray}
where we have dropped terms of ${\cal O} (q^4)$ and higher.
(Recall that these expansions are being used not to evaluate
${\cal E}_V$ but to determine ${\cal E}_{ctc}$.)
Upon discretization of $p$ and $p'$,
\begin{equation} \tilde S (q)\rightarrow {a\over 2} f_n S_n \end{equation}
where
\begin{equation}     f_n=\cases{2&for $n=0$; \cr
		     1&otherwise \cr} \end{equation}
and the coefficients $S_n \ (=S_{-n} )$ are those defined in
the truncated Fourier series expansion of the scalar potential, $S(x)$,
appearing in Eqn.~\ref{bm} .  Then
\begin{eqnarray} {\cal E}^{(2)}_V (x)={1\over 2}S(x)\ \rho^{(1)}_V(x)
	\simeq{1\over 2} S(x)\cdot {1\over a} \sum_{i,j=-N+1}^N &&
	{{\tilde f_{ij} \tilde S_{ij}}\over {2}}\ \cos q_{ij} x \nonumber \\
	&&\times\bigl[{\cal F}_0 (k_{ij}) +
	  {\cal F}_2 (k_{ij}) q^2_{ij} \bigr] \label{cp} \end{eqnarray}
where $\tilde f_{ij}\equiv f_{i-j}$, $\tilde S_{ij}\equiv S_{i-j}$,
$q_{ij}\equiv 2(i-j)\pi/a$ and $k_{ij}\equiv (i+j-1)\pi/a$.  Observing
that
\begin{equation} \sum_{n=-2N+1}^{2N-1} {{ f_n S_n}\over{2}}\
	\cos q_n x=\sum_{n=0}^{2N-1} S_n \cos q_n x \end{equation}
which is the truncated expansion of the scalar potential, we identify
\begin{equation} {{\alpha_2}\over{2!}}\ S^2(x)\leftrightarrow {1\over 2} S(x)
	\cdot {1\over a}\sum_{ij} {{\tilde f_{ij} \tilde S_{ij}}
	\over{2}}\ \cos q_{ij} x \ {\cal F}_0(k_{ij}) \label{cq} \end{equation}
and
\begin{equation} {1\over 2}\ \xi\ \partial S\cdot\partial S =
	-{1\over 2}\ \xi\ [S']^2
	\rightarrow{1\over 2}\ \xi\ S S'' \leftrightarrow
	  {1\over 2}\ S(x)\ \biggl[\sum_{ij} {{\tilde f_{ij}
	  \tilde S_{ij}}\over{2}}\ \cos q_{ij} x\ {\cal F}_2 (k_{ij})
	  q^2_{ij} \biggr] \label{crr} \end{equation}
\narrowtext
where we have integrated by parts and
dropped surface terms. 
These identifications require some discussion because at first glance
the quantities on the r.h.s.~do not appear to factorize into the
forms given on the l.h.s.  However, because of the definitions of
$q_{ij}$ and $k_{ij}$, the former depending on $i-j$ and the latter on
$i+j$, the summations over $i$ and $j$ may be expressed approximately as
independent sums over $q_{ij}$ and $k_{ij}$ in which case the required
factorizations follow immediately.  However, near the end of the
truncated basis, the sums over $q_{ij}$ and $k_{ij}$ become correlated
and we must therefore use the full expressions given above.  Although we
have not demonstrated it rigorously, it seems very likely that the
counter term subtractions we employ can be justified formally by
taking care to define the original Lagrangian in terms of the degrees of
freedom specified by the truncated basis.

	Next
\widetext
\begin{equation}
	{\cal E}^{(3)}_V (x)\rightarrow {1\over 3}\ S(x)\ \rho^{(2)}_V (x)
\end{equation}
where, upon defining $q=p_2-p_1$, $q'=p_3-p_2$ and $k=(p_1+p_3)/2$,
\begin{eqnarray} \rho^{(2)}_V (x)=\int_{-\infty}^{+\infty}
	{{d\omega}\over{2\pi}}&&\
	e^{+i\omega 0^+} \int_{-\infty}^{+\infty} dy\
	\int_{-\infty}^{+\infty} dy' \nonumber \\
	&&\times {\rm Tr}\ \bigl[G_0(x-y;\omega)\
	S(y)\ G_0(y-y';\omega)\ S(y')\ G_0(y'-x;\omega) \bigr] \nonumber \\
	&&\rightarrow \int_{-\infty}^{+\infty} {{dp_1}\over{2\pi}}\
	  \int_{-\infty}^{+\infty} {{dp_2}\over{2\pi}}\
	  \int_{-\infty}^{+\infty} {{dp_3}\over{2\pi}}\
	  \tilde S(q)\ \tilde S(q')\ \cos (q+q') x \nonumber \\
	&&\qquad\qquad\times\theta(\Lambda-|p_1|)\
	  \theta(\Lambda-|p_2)\ \theta(\Lambda-|p_3|)
	  \bigl[{\cal G}_0 (k) + {\cal O}(q^2) \bigr] \label{cs}
\end{eqnarray}
where ${\cal G}_0 (k) 
	=-\partial{^3 E_k}/\partial{m^3}$.
After dropping the ${\cal O} (q^2)$ and higher terms and then
discretizing, we identify
\begin{eqnarray} {{\alpha_3}\over{3!}}\ S^3(x)\leftrightarrow {1\over 3}
	\ S(x)\cdot {1\over a} \sum_{i,j=-N+1}^N \ \sum_{m=-N+1}^N
	&& {{\tilde f_{im} \tilde S_{im}}\over{2}}\
	  {{\tilde f_{mj} \tilde S_{mj}}\over{2}} \nonumber \\
	&&  \qquad\times\cos q_{ij} x\ {\cal G}_0 (k_{ij}).
	  \label{ct} \end{eqnarray}
A similar procedure gives
\begin{eqnarray} {{\alpha_4}\over{4!}}\ S^4(x)\leftrightarrow {1\over 4}
	\ S(x)\cdot {1\over a} \sum_{i,j=-N+1}^N\qquad\sum_{m,m'=-N+1}^N
	&& {{\tilde f_{im}  \tilde S_{im}}\over{2}}\
	  {{\tilde f_{mm'} \tilde S_{mm'}}\over{2}}\
	  {{\tilde f_{m'j} \tilde S_{m'j}}\over{2}} \nonumber \\
	&&  \qquad\times\cos q_{ij} x\ {\cal H}_0 (k_{ij})
	  \label{cu} \end{eqnarray}
where ${\cal H}_0 (k) 
	=-\partial{^4 E_k}/\partial{m^4}$.  Observing that contributions
to the {\it energy} are found by
\begin{equation}
	E^{(n)}_V=\int_{-a/2}^{+a/2} dx\ {\cal E}^{(n)}_V (x), \end{equation}
we determine that the subtraction required for renormalization of the
1-fermion loop energy is
\begin{eqnarray} \sum_{i=-N+1}^N\biggl[-E_i-S_0{m\over{E_i}}\biggr] +
	\sum_{i,j=-N+1}^N &\biggl\{& {1\over 2}\bigl[{\cal S}_{ij}^{(1)}
	\bigr]^2 \bigl[ {\cal F}_0 (k_{ij}) + {\cal F}_2 (k_{ij})\
	q_{ij}^2\bigr] \nonumber \\
	&+& {1\over 3} {\cal S}_{ij}^{(1)} {\cal S}_{ij}^{(2)}
	  {\cal G}_0 (k_{ij}) + {1\over 4} {\cal S}_{ij}^{(1)}
	  {\cal S}_{ij}^{(3)} {\cal H}_0 (k_{ij}) \biggr\}
	\label{cv} \end{eqnarray}
while the density subtraction is
\begin{eqnarray}  {1\over a} \sum_{i=-N+1}^N\biggl[-{m\over{E_i}}\biggr] +
	{1\over a} \sum_{i,j=-N+1}^N &\biggl\{& {\cal S}_{ij}^{(1)}\
	\bigl[ {\cal F}_0 (k_{ij}) + {\cal F}_2 (k_{ij})\
	q_{ij}^2\bigr] \nonumber \\
	&+& {\cal S}_{ij}^{(2)} {\cal G}_0 (k_{ij}) +
	  {\cal S}_{ij}^{(3)} {\cal H}_0 (k_{ij}) \biggr\} \cos q_{ij} x.
	  \label{cw} \end{eqnarray}
In these expressions,
\begin{equation} {\cal S}_{ij}^{(1)}\equiv {{\tilde f_{ij}
	\tilde S_{ij}}\over{2}},
	\qquad {\cal S}_{ij}^{(2)}\equiv \sum_{m=-N+1}^N
	{\cal S}_{im}^{(1)} {\cal S}_{mj}^{(1)},
	 \qquad\qquad {\cal S}_{ij}^{(3)}\equiv \sum_{m=-N+1}^N
	  {\cal S}_{im}^{(1)} {\cal S}_{mj}^{(2)} \end{equation}
and
\begin{eqnarray} {\cal F}_0 (k)=&&-\partder{^2E_k}{m^2},\quad {\cal F}_2 (k)=
	{{5k^2m^2}\over{8E_k^7}}, \nonumber \\
	{\cal G}_0 (k)=&&-\partder{^3E_k}{m^3},\quad {\cal H}_0 (k)=
	  -\partder{^4E_k}{m^4}. \end{eqnarray}

\narrowtext
	We now observe that these subtractions are {\it inconsistent}
with the discrete chiral symmetry of the original Lagrangian (Eqn.~\ref{ba}
).  A counter term Lagrangian possessing this symmetry is
\begin{equation} -{\cal L}_{ctc}\rightarrow {1\over 2}\ \xi\ \partial\phi\cdot
	\partial\phi + a_1(\phi^2-f^2) +a_2(\phi^2-f^2)^2 \label{cx}
\end{equation}
(compare with Eqn.~\ref{cf} ).  Using
\begin{equation} \phi^2-f^2={1\over{g^2}} S\cdot(2m+S) \end{equation}
we may write (redefining $a_1$ and $a_2$)
\begin{equation} -{\cal L}_{ctc}={1\over 2}\ \xi\ \partial S \cdot\partial S +
	a_1 S\cdot(2m+S) + a_2 S^2\ (2m+S)^2 \end{equation}
Evidently the derivative counter term contribution is unaffected by the
imposition of chiral symmetry.  The remaining coefficients are fixed by
expanding (assuming $\phi$ is uniform)
\widetext
\begin{eqnarray}  E_i \bigl[ (g\phi)^2 \bigr]\equiv&&\sqrt{k_i^2+(g\phi)^2}
	\nonumber \\
	&&\simeq E_i^0 + g^2(\phi^2-f^2)\partder{E_i^0}{(gf)^2}
	  +{1\over{2!}} g^4(\phi^2-f^2)^2\partder{^2 E_i^0}
	  {[(gf)^2]^2} \label{cxx} \end{eqnarray}
where  $E_i^0\equiv\sqrt{k_i^2+(gf)^2}$.  Thus, again dropping the
superscript on $E_i^0$, the non-derivative chiral subtraction for the
energy is
\begin{eqnarray} -\biggl\{ E_i + S\cdot 2m\partder{E_i}{m^2}&& +S^2\biggl[
	\partder{E_i}{m^2} +{1\over 2}\cdot 4m^2\partder{^2E_i}
	{[m^2]^2}
	\biggr] \nonumber \\
	&& + S^3\cdot {1\over 2}\cdot 4m \partder{^2E_i}{[m^2]^2}
	  +S^4\cdot {1\over 2} \partder{^2E_i}{[m^2]^2} \biggr\} .
	\label{cy} \end{eqnarray}
Since $2m\ \partial E_i /\partial m^2 = \partial E_i /\partial m$, the
energy and scalar density subtractions of Eqn.~\ref{cv}\ and \ref{cw} ,
respectively, are made chirally symmetric by redefining
\begin{eqnarray} {1\over 2}{\cal F}_0 (k)\rightarrow&&
	{1\over 2}{\cal F}_0^{(0)}(k) + {1\over 2}{\cal F}_0^{(1)}(k)
	=-\partder{E_k}{m^2} - 2m^2\partder{^2E_k}{[m^2]^2}, \nonumber \\
	&&{1\over 3}{\cal G}_0 (k) = -2m\partder{^2E_k}{[m^2]^2},\quad
	 {1\over 4}{\cal H}_0 (k) = -{1\over 2}\partder{^2E_k}{[m^2]^2}.
	 \label{cz} \end{eqnarray}
\narrowtext
The chirally renormalized 1-fermion loop {\it energy} is then
\begin{equation} E_{loop}=\sum_{n'=1}^N\sum_\pi E^{(\pi,-)}_{n'}
	-\sum_{n=0}^4 E^{(n)}_V \label{caa} \end{equation}
while the {\it density} is
\begin{equation} \rho_{loop} (x) = \sum_{n'=1}^N \sum_\pi
	\bar\psi^{(\pi,-)}_{n'} (x)
	\psi^{(\pi,-)}_{n'} (x) - \sum_{n=0}^3 \rho^{(n)}_V \label{cab}
\end{equation}
where $E^{(\pi,-)}_{n'}$ and $\psi^{(\pi,-)}_{n'}$ are the negative energy
eigenvalues and eigenspinors determined by solving the Dirac equation
(Eqn.~\ref{bg} ) and where $\pi$ refers to the parity of the solution.

	It is straightforward to re-express the renormalized quantities
given above via a derivative expansion (DE).  While a DE is inadequate to
describe the entire sea contribution, we may use a modified DE to
account for effects due to those components of the sea lying {\it outside}
our truncated basis.  Details of this DE are given in Ref.~\cite{fer} .  The
calculations to be presented below always employ the renormalized
quantities of Eqns.~\ref{caa}\ and \ref{cab}\ plus DE contributions from
{\it outside} the truncated basis.

	We now note that the chiral renormalization scheme just outlined
differs from those of earlier studies\cite{candl,wass} where,
in effect, it was assumed that
\begin{equation} {\cal F}_0^{(1)}={\cal F}_2={\cal G}_0={\cal H}_0=0.
	\label{da}
\end{equation}
In other words, these earlier approaches include only the first {\it two}
terms on the
r.h.s. of Eqn.~\ref{cxx}\ and also drop the derivative counter term.  In 1+1
dimensions the resulting energy and density are finite.  We refer to
this as ``minimal'' chiral renormalization as opposed to the ``full''
renormalization presented above.  Note that ${\cal F}^{(1)}_0$,
${\cal G}_0$ and ${\cal H}_0$ all originate with the last term on
the r.h.s. of the expansion of Eqn.~\ref{cxx}\ and must either {\it all}
be dropped or {\it all} be included to preserve chiral symmetry.  We
also note that some analogue of ``full'' renormalization is {\it
required} in 1+3 dimensions to yield finite 1-loop contributions.

	The physics implied by the Lagrangian of Eqn.~\ref{ba}\ is of course
ultimately {\it independent} of the renormalization scheme.  For any set
of parameters $\lambda$, $f^2$, and $g$ using the ``full'' scheme, there
is another set which will yield the same physics in the ``minimal''
scheme.  In most of what follows, we will employ the full scheme since
(1) the ``extra'' counter terms are required in three spatial dimensions
and (2) the 1-fermion loop contributions to the $\sigma$ propagator
vanish at the renormalization point (namely, $\sigma = q^2 = 0$.)

\section{RESULTS}

	We test the numerics of our calculations by comparing with the
published results of Campbell and Liao\cite{candl} and of
Wasson,\cite{wass} both of which employ minimal renormalization.
For this renormalization scheme, kink solutions with $\lambda/g^2=2$
constitute a special case in that the 1-fermion
loop contribution to the scalar density vanishes when $\phi (x)$
is given by the classical kink solution, Eqn.~\ref{bxd} .  Hence, this
kink solution is also {\it self-consistent} (with or without any fermions
in the lowest positive energy level).
Campbell and Liao treat only this case and find an analytic
expression for the total kink energy, namely, their Eqn.~4.15.  Their
result implies $E_{loop} =g/\pi \simeq 0.318310 g$.
Table \ref{taba}\ shows our evaluations of this quantity for various basis
sizes ($N_{order}$) and iteration numbers ($N_{iter}$).  (Also shown
in Table \ref{taba}\ are the maximum momentum in the basis $(\Lambda/g)$,
and the percentage DE contribution to the energy.)  We evidently agree
well with Campbell and Liao.  In fact, the $N_{order}=100$ calculation
-- when retaining more significant figures than are displayed in
Table \ref{taba}\ -- differs from analytic result by less than 0.004\% .
Wasson reports that the 1-fermion loop energy in this case (designated
as ``G=1'' in Table 4 of Ref.~\cite{wass} ) is $0.225\sqrt{\lambda} g=0.318
g$ and, of course, we agree with him.  He also gives non-iterated (or
perturbative) as well as self-consistent results for {\it other} values of
$\lambda/g^2$.  When using adequate basis sizes ($N_{order}\simeq 25$ or
larger) and a sufficient number of iterations (typically $N_{iter}\geq
10$), we agree with {\it all} of Wasson's published results.
Table~\ref{taba}\ shows this agreement for $\lambda /g^2=0.5$ as well as for
the special case of $\lambda /g^2=2$.  For the former coupling we
reproduce Wasson's non-iterated, or ``perturbative'' 1-loop energy of
$0.348 g$ and his iterated or ``self-consistent'' value of $0.338 g$
exactly.  These examples demonstrate the accuracy and efficiency of our
calculational method.  Table~\ref{taba}\ also reveals that DE contributions
to the 1-loop energy from outside the truncated basis are always much
less than 1\% (except for the very smallest basis size of
$N_{order}=8$).


	Table \ref{tabb}\ presents similar results, but now utilizing the
full renormalization discussed above.  We first note that the
1-loop energies are roughly two-thirds of those found with the minimal
scheme.  It also appears that the classical kink solution is
self-consistent in the full calculations for $\lambda/g^2=0.5$ rather
than $\lambda/g^2=2$ as before.  Furthermore, for those situations where
the classical kink is {\it not} self-consistent, the
self-consistent 1-loop energies typically differ from the perturbative
values by less than 0.5\% while such differences
are $\sim$ 3\% with minimal renormalization.   Finally, DE contributions
from outside the
basis are roughly two orders of magnitude smaller when computed with
full renormalization.  Remaining sensitivities to basis size are tiny.
It is thus apparent that the full chiral renormalization -- in effect --
{\it cuts off} 1-loop contributions from the fermion sea at quite small
values of momentum.  For example, we see (Table \ref{tabb} ) that, for
$\lambda/g^2=2$, less than $1/2$ \% of the 1-loop energy is due to levels
in the sea with momentum greater than $\sim 2g$, {\it i.e.}, $\sim2
\times$ the asymptotic fermion mass, $m=-gf$.  Such a drastic
(effective) cutoff is surprising in light of the huge changes in the
fermion effective mass implied by the kink solution, namely
\begin{equation} 0\leq\biggl({{m^*}\over{m}}\biggr)^2=
	\biggl({\phi\over f}\biggr)^2
	\leq 1. \end{equation}
Put another way, fermions whose mass-squared is $g^2 f^2$ far from the kink
are subject to interactions so strong that their mass {\it vanishes} at
the center of the kink.  Such interactions could be expected to polarize
the negative energy sea very significantly and would seem to suggest
much larger 1-loop contributions from deep in the sea than are observed,
especially with full renormalization.  (We note in this
regard that, for the uniform system, the 1-fermion loop energy density
possesses a logarithmic singularity at $m^*=0$.)

	We now turn to a comparison of ``kink'' {\it versus} ``shallow bag''
solutions.  It is assumed in {\it both} cases that the lowest positive
energy fermion level is fully occupied, {\it i.e.}, this state contains
$N_{sea}$ fermions where $N_{sea}$ is the fermion multiplicity owing to,
{\it e.g.}, color, flavor, etc.
(Note that $N_{sea}=1$ was assumed throughout the preceding discussion
in order to compare with Wasson.)  For the kink solutions the lowest positive
energy fermion state ({\it and} the highest negative energy state) has zero
total energy and yields a vanishing scalar density.  Hence
occupancy of this state has {\it no effect} on either the 1-loop energy or
the scalar field configuration.  In contrast, the positive energy
fermions are {\it essential} for the shallow bag solutions since the
associated scalar fields would decay to the trivial uniform
configuration ($\phi^2=f^2$) in their absence.


	Properties of these solutions obtained using full
renormalization are displayed in Table \ref{tabd}.  Also,
Figure~\ref{figb}\ shows vector and scalar densities as well as
scalar potentials for self-consistent full kink and shallow
bag solutions with $\lambda/g^2=0.5$ and $N_{sea}=6$.  Figure~\ref{figd}\
presents the same quantities but for $N_{sea}=18$.  Note that
the scalar potential is $S(x)$ as defined in Eqn.~\ref{bff} .  Also, the vector
and scalar densities are
\begin{equation} \rho_0 (x)=\sum_{occ'd}\bar\psi^{(+)}_i
	\gamma^0 \psi^{(+)}_i \end{equation}
and
\begin{equation} \rho_s (x)=\sum_{occ'd}\bar\psi^{(+)}_i \psi^{(+)}_i +
	\rho_{loop} (x), \end{equation}
respectively, where the sums are over positive energy occupied levels
and $\rho_{loop}$ is the renormalized sea contribution to the scalar
density given in Eqn.~\ref{cab} .  In the present case, with scalar potentials
only, the sea makes no contribution to the vector density.\cite{fer}
Because of the
symmetries built into our free basis (Eqn.~\ref{bj} ) which ensure, for
example, that all densities are symmetric about $x=0$, ``kinks'' occur
in pairs located symmetrically about the origin.  (See Figures~\ref{figb}\ and
\ref{figd}.)
Furthermore, the vector ({\it i.e.},
probability) density for the lowest positive energy fermion state
associated with the kinks is strongly localized at the center of each
kink where $S(x)=-g$ and $m^*$ vanishes.
Hence, each kink contains $N_{sea}/2$
fermions.  In contrast, the shallow bag solutions are localized near the
origin and each contains $N_{sea}$ fermions.  For this reason, we
compare quantities for {\it two} kinks with results for one
shallow bag in Tables \ref{tabc}\ and \ref{tabd}.
We note also (Figures~\ref{figb}\ and \ref{figd} ) that the fully
renormalized kink scalar densities for $\lambda/g^2=0.5$ {\it vanish};
in fact, the valence and sea contributions vanish separately.  The same
phenomenon is, of course, observed for $\lambda /g^2=2$ with minimal
renormalization.

	Armed with analytic results for $\lambda /g^2=2$, Campbell and
Liao\cite{candl} determined that, in this special case, shallow bags
are more bound than kinks for $N_{sea}< 4\pi$ while for $N_{sea}> 4\pi$,
the situation is reversed.  Our numerical results are consistent with
this finding. The present calculations show that, for $N_{sea}< 4\pi$,
self-consistent kink solutions can be found which are {\it metastable}
with respect to the shallow bag while, for $N_{sea}> 4\pi$, the shallow
bag is {\it unstable} and only the kink can be self-consistent.  As
$\lambda /g^2$ is reduced from the special value of 2, the value of
$N_{sea}$ for which kinks and shallow bags become degenerate appears to
increase from $4\pi$.  Since full renormalization reduces 1-loop
contributions, the disappearance of the shallow bag as a stable solution
appears at a larger value of $N_{sea}$ for a given parameter set; at
$N_{sea}=18$, Table~\ref{tabd}\ shows that the fully renormalized shallow bag
is still more bound than the corresponding kink.
We have done additional fully renormalized calculations for $N_{sea}=18$
which suggest
that, for this multiplicity, kink {\it versus} shallow bag degeneracy
occurs for some value of $\lambda /g^2$ between 3 and 4.

	It is interesting to inspect the shallow bag vector densities for
$N_{sea}=6$
(Figure~\ref{figb} ) and $18$ (Figure~\ref{figd} ) and to see that, for
$\lambda /g^2=0.5$ when the shallow bag is always more bound, larger
values of $N_{sea}$ result in shallow bags which look increasingly like
slightly overlapping kinks.
This leads to the intriguing notion that the shallow bag can -- for
large values of $N_{sea}$ but where the shallow bag is still
energetically favored -- be viewed as a weakly bound pair of kinks.

	Although the principal aim of the present paper is (1) to
demonstrate the utility of our calculational method and (2) to examine
the relative stability of standard kink and shallow bag solutions, it
seems of interest to examine fully renormalized
kink and shallow bag solutions for
``baryons'' consisting of {\it three} fermions in the lowest positive
energy state with $N_{sea}=6$.
The kink solutions whose properties are summarized in Tables
\ref{tabc}\ and \ref{tabd}\ are of course just {\it two} such baryons.
Further,
the kink densities shown in Figure~\ref{figb}\ represent {\it
one} kink baryon.  New shallow
bag solutions are required.  Properties of these shallow bag baryons
are displayed in Table \ref{tabe}\ and the associated densities and scalar
potentials are displayed in Figure~\ref{fige} .
For $\lambda/g^2=2$ and minimal renormalization,
the shallow bag solutions are
more bound than the kink configurations.  However, as $\lambda/g^2$ is
reduced and the relative scalar-fermion coupling goes up, the situation
is reversed.  With $\lambda/g^2=0.5$ and minimal renormalization, the
binding energy/fermion for the kink is more than twice that of the
shallow bag.  When using the full renormalization, the kink binding
is always greater than that of the shallow bag.  Comparison of the kink
vector densities shown in Figure~\ref{figb}\ with the shallow bag
vector densities of Figure~\ref{fige}\ show that the former have widths
which are roughly $1/2$ those of the latter as might be expected from
the binding energy differences.
For the most strongly bound fully
renormalized kink solution, the binding energy/fermion is
nearly $1/3$ the asymptotic fermion mass.  We have made no effort to
factor out center-of-mass effects\cite{candl,col} from our
solitonic solutions and are therefore at some remove from ascribing
physical significance to them even after extrapolating to 1+3 dimensions.
However, it is interesting to note that kink ``baryons'' are typically
more bound than their shallow bag counterparts even with 1-fermion loop
effects.

	All results quoted so far assume $f^2=1$.  The kink solutions
are dominated by scalar self-interactions.  Hence, recalling all
energies are expressed in terms of $m=|gf|$, it is not suprizing that,
to a good approximation, $E_{loop}/N_{sea}$ is independent of all parameter
variation while $E_{meson}\propto \sqrt{\lambda /g^2}\cdot f^2$.
Therefore any change in $f$ can be closely approximated by varying
$\lambda /g^2$ and no qualitatively new features emerge upon changing
$f^2$ from unity.  For ``shallow'' shallow bag solutions, the situation
is different since they are dominated by scalar-fermion interactions.
To a first approximation, the total energy of these solutions is
independent of $f^2$.  At a more detailed level, since $E_{meson}
\propto f^2$, the total energy grows slowly as $f^2$ is increased.  Of
course, as $N_{sea}$ is increased and the shallow bag solutions approach
the kink solutions, the scalar-scalar interactions become more important
in the former.  For the present purposes, the situation is perhaps best
summarized by observing that -- with full renormalization -- variation
of $f^2$ does not alter the fact that standard shallow bag solutions are
always more bound than kinks even though differences are often very small.

\section{SUMMARY AND CONCLUSIONS}

	We have examined solutions to the 1+1 dimensional linear sigma
model with discrete chiral symmetry focussing on ``kink'' {\it versus}
``shallow bag'' solutions.  One fermion loop contributions have been
treated {\it exactly} via a new method\cite{fer} which proves to be
accurate and efficient.  Such a treatment is desirable in light of the
very strong interactions to which the fermions are subject, especially in
the case of the kink configurations.  Our method permits us to go beyond
the ``special cases'' considered in early studies.\cite{candl}  In
examining shallow bag solutions, we also go beyond the recent work of
Wasson.\cite{wass}  We furthermore investigate a ``full''
renormalization scheme which is different from those of previous works
where only those subtractions required to eliminate divergences were
performed and to which we refer as ``minimal''.  Physically, the full
renormalization prescription differs from the minimal in that it ensures
that, at the renormalization point, the scalar
propagator dressed by 1-fermion loop effects coincides with its
classical counterpart.  One-loop effects are generally suppressed by full
renormalization relative to the minimal results.  Finally, we note that
some analogue to our full
renormalization is required in 1+3 dimensions to yield finite 1-fermion
loop contributions.

	Specific calculations show that our method reproduces the
results of Campbell and Liao\cite{candl} and of
Wasson\cite{wass}
exactly even with modest basis sizes.  We find that, when
employing full renormalization, the case of $\lambda/g^2=0.5$ appears to
be special in that the 1-loop contribution to the scalar density vanishes
implying that the classical kink solution is also self-consistent.  With
minimal renormalization, a similar special case is known to
exist\cite{candl} for $\lambda/g^2=2$.  We find that, overall, the
effects of self-consistency are considerably less in the full scheme
than in the minimal one.  Further, full renormalization effectively {\it
cuts off} the 1-loop contributions so that only fermion states {\it
very} near the top of the negative energy sea are of any consequence.
For example, less than 1/2\% of the 1-loop energy in the full calculations
is due to states with momenta greater than 2 fermion masses.

	We have carefully compared kink and shallow bag solutions for
which the lowest positive energy fermion states are filled using a
variety of couplings and fermion multiplicities.  Our results show that,
for a given value of $\lambda /g^2$,
distinct self-consistent kink and shallow bag solutions exist for small
values of $N_{sea}$, the latter being more bound.  However, beyond some
critical value, {\it e.g.}, $N_{sea}=4\pi$ for $\lambda /g^2 =2$ and minimal
renormalization\cite{candl}), only kink solutions are
self-consistent.  When full renormalization is employed, the critical
value of $N_{sea}$ is greater than for the minimal results.

	Although the present work emphasizes (1) the accuracy and
efficiency of the calculational method and (2) the relative stability
the standard kink and shallow bag solutions, we have also looked at
kink and shallow bag ``baryons''.  Here, we
have assumed an overall fermion multiplicity of 6 and that each
localized soliton contains 3 positive energy fermions.  We find that,
with full renormalization, kink baryons are always more bound
than those of the shallow bag.  For $\lambda/g^2=0.5$, the fully renormalized
binding energy per
fermion is $\sim$ 1/3 of the fermion mass.  While it is questionable to
attach much physical significance to our ``baryons'', it is interesting
that kink ``baryons'' are typically more bound than their shallow bag
counterparts even when 1-fermion loop effects are included.

	Several extensions of the present study are either in progress
or being contemplated.  For example, we are in the process of modifying
our calculational method to treat 1-{\it scalar} loop contributions --
which are attractive in contrast to the repulsive fermion loops -- (see
Wasson and Koonin, Ref.~\cite{wass} ) and plan to find solutions for which
{\it all} 1-loop effects are treated in a fully self-consistent fashion.
Also planned is the extension to the full linear sigma model containing
a pseudoscalar field and possessing {\it continuous} chiral symmetry.  More
interesting, and certainly more challenging, is the extension
to 1+3 dimensions.  This is not a trivial undertaking, but preliminary
results for the scalar-only problem are encouraging and an
approach for fermions has been outlined.  The efficiency of our
calculational method is especially encouraging as we contemplate going to
higher dimensions.



\section{Acknowledgements} This work supported in part by the U.S.D.O.E.

\vfill
\eject
\newpage
%
%
%
%

%
%
%
\newpage
\figure{ Densities and scalar potentials for the $\lambda/g^2=0.5$,
	$f^2=1$ and $N_{sea}=6$ kink and shallow bag solutions
	of Table~\ref{tabd}, {\it i.e.},
	full renormalization is employed. Lengths are in units of $1/g$
	while densities and energies are in units of $|gf|$.  All quantities
	displayed are {\it even} functions of $x$. \label{figb}}
\figure{ Same as Fig.~\ref{figb}\ but with $N_{sea}=18$. \label{figd}}
\figure{ Densities and scalar potentials for shallow bag ``baryons''
	whose properties are presented along with kink counterparts in
	Table~\ref{tabe}.  Units are the same as in Figures~\ref{figb}\
	and \ref{figd}\
	but note that the scales are  expanded by a factor of 2 here.
	The corresponding kink ``baryons''  appear in Figure~\ref{figb}.
	See text for discussion. \label{fige}}
%
%
%
%
\newpage
\begin{table}
\caption{ Properties of ``sea-only'' kink solutions with minimal
	renormalization scheme.  Assumes $N_{sea}=1$, $f^2=1$. $E_{loop}$
	is the renormalized energy contribution due to the sea, {\it i.e.},
	$\tilde E_V$ as defined in Eqn.~\ref{caa}\ plus DE contributions from
	outside the basis.  Also shown is the DE percentage of the 1-loop
	energy. }
\begin{tabular}{cccccc}
\multicolumn{1}{c}{$\lambda/g^2$} &\multicolumn{1}{c}{$N_{order}$}
&\multicolumn{1}{c}{$\Lambda/g$} &\multicolumn{1}{c}{$N_{iter}$}
&\multicolumn{1}{c}{$E_{loop}/g$} &\multicolumn{1}{c}{\% DE} \\
\tableline
2.0 &   8 &  1.98 &   1 &  0.3195 & 1.66 \\
    &  25 &  6.47 &   1 &  0.3185 & 0.19 \\
    &\    &\      &  10 &  0.3185 & 0.19 \\
    &  50 & 13.07 &   1 &  0.3183 & 0.05 \\
    &\    &\      &  10 &  0.3183 & 0.05 \\
    & 100 & 26.28 &   1 &  0.3183 & 0.01 \\
\multicolumn{6}{c}{\ } \\
0.5 &   8 &  1.98 &   1 &  0.3507 & 3.03 \\
    &  15 &  3.83 &   1 &  0.3484 & 0.94 \\
    &\    &\      &   4 &  0.3391 & 0.87 \\
    &\    &\      &  15 &  0.3388 & 0.86 \\
    &  25 &  6.47 &   1 &  0.3480 & 0.35 \\
    &\    &\      &  15 &  0.3383 & 0.32 \\
    &  50 & 13.07 &   1 &  0.3479 & 0.09 \\
    &\    &\      &  15 &  0.3382 & 0.08 \\
\end{tabular}
\label{taba}
\end{table}

\newpage
\begin{table}
\caption{ Same as Table \ref{taba} , but with full renormalization. }
\begin{tabular}{cccccc}
\multicolumn{1}{c}{$\lambda/g^2$} &\multicolumn{1}{c}{$N_{order}$}
&\multicolumn{1}{c}{$\Lambda/g$} &\multicolumn{1}{c}{$N_{iter}$}
&\multicolumn{1}{c}{$E_{loop}/g$} &\multicolumn{1}{c}{\% DE} \\
\tableline
0.5 &   8 &  1.98 &   1 &  0.2335 & 0.09 \\
\   &  25 &  6.47 &   1 &  0.2330 & 0.00 \\
\   &\    &\      &  10 &  0.2330 & 0.00 \\
\   &  50 & 13.07 &   1 &  0.2330 & 0.00 \\
\   &\    &\      &  10 &  0.2330 & 0.00 \\
\   & 100 & 26.28 &   1 &  0.2330 & 0.00 \\
\multicolumn{6}{c}{\ } \\
2.0 &   8 &  1.98 &   1 &  0.2466 & 0.18  \\
\   &  15 &  3.83 &   1 &  0.2477 & 0.008 \\
\   &\    &\      &   4 &  0.2469 & 0.008 \\
\   &\    &\      &  15 &  0.2468 & 0.008 \\
\   &  25 &  6.47 &   1 &  0.2476 & 0.00  \\
\   &\    &\      &  15 &  0.2467 & 0.00  \\
\   &  50 & 13.07 &   1 &  0.2476 & 0.00  \\
\   &\    &\      &  15 &  0.2467 & 0.00  \\
\end{tabular}
\label{tabb}
\end{table}

\widetext
\newpage
\begin{table}
\caption{ Kink {\it versus} Shallow Bag Solutions with minimal
	renormalization.  Assumes lowest positive energy fermion
	level
	filled and $f^2=1$.  $E_{loop}$ is the same as for Table \ref{taba}.
	$E_{meson}$ is the energy due to the scalar field and $E_{val}$
	is that for the positive energy fermions.  BE/fermion$=
	(E_{tot}-N_{val}\ g)/N_{val}$ where $N_{val}$ is the number of
	positive energy fermions.  $S(0)$ is the shallow bag scalar
	potential at $x=0$. }
\begin{tabular}{cccccccc}
\multicolumn{1}{c}{$N_{sea}$} & \multicolumn{1}{c}{$\lambda/g^2$}
&\multicolumn{1}{c}{$E_{meson}$} & \multicolumn{1}{c}{$E_{loop}$}
&\multicolumn{1}{c}{$E_{val}$} & \multicolumn{1}{c}{$E_{tot}$}
&\multicolumn{1}{c}{BE/fermion} & \multicolumn{1}{c}{$S(0)$} \\
\multicolumn{8}{l}{ Kink Solution } \\
 3 & 2.0 & 2.6667 & 1.9090 & 0      & 4.5766 & +0.5256 &\        \\
\  & 1.0 & 1.8932 & 1.9306 & 0      & 3.8240 & +0.2756 &\        \\
\  & 0.5 & 1.3734 & 1.9780 & 0      & 3.3516 & +0.1172 &\        \\
\multicolumn{8}{c}{\ } \\
 6 & 2.0 & 2.6667 & 3.8198 & 0      & 6.4866 & +0.0811 &\        \\
\  & 1.0 & 1.9043 & 3.8461 & 0      & 5.7505 & -0.0416 &\        \\
\  & 0.5 & 1.4158 & 3.8965 & 0      & 5.3125 & -0.1146 &\        \\
\multicolumn{8}{c}{\ } \\
18 & 2.0 & 2.6667 &11.4594 & 0      &14.1265 & -0.2152 &\        \\
\  & 1.0 & 1.9346 &11.4842 & 0      &13.4195 & -0.2545 &\        \\
\  & 0.5 & 1.5090 &11.5224 & 0      &13.0320 & -0.2760 &\        \\
\multicolumn{8}{c}{\ } \\
\multicolumn{8}{l}{ Shallow Bag Solution } \\
 3 & 2.0 & 0.0861 & 0.0214 & 2.8445 & 2.9520 & -0.0160 & -0.1045 \\
\  & 1.0 & 0.2004 & 0.1029 & 2.5579 & 2.8613 & -0.0462 & -0.2786 \\
\  & 0.5 & 0.2944 & 0.2956 & 2.1238 & 2.7138 & -0.0954 & -0.5089 \\
\multicolumn{8}{c}{\ } \\
 6 & 2.0 & 0.4740 & 0.2615 & 4.9618 & 5.6973 & -0.0505 & -0.3461 \\
\  & 1.0 & 0.7565 & 0.8960 & 3.6653 & 5.3204 & -0.1133 & -0.7223 \\
\  & 0.5 & 0.6858 & 1.4689 & 2.8162 & 4.9708 & -0.1715 & -0.9347 \\
\multicolumn{8}{c}{\ } \\
18 & 2.0 & 2.6649 &11.0840 & 0.3783 &14.1268 & -0.2152 & -1.9582 \\
\  & 1.0 & 1.8959 &10.3681 & 1.1435 &13.4075 & -0.2551 & -1.8366 \\
\  & 0.5 & 1.3286 & 9.0000 & 2.6562 &12.9848 & -0.2786 & -1.6217 \\
\end{tabular}
\label{tabc}
\end{table}

\newpage
\begin{table}
\caption{ Same as Table \ref{tabc}\ but with full renormalization. }
\begin{tabular}{cccccccc}
\multicolumn{1}{c}{$N_{sea}$} & \multicolumn{1}{c}{$\lambda/g^2$}
&\multicolumn{1}{c}{$E_{meson}$} & \multicolumn{1}{c}{$E_{loop}$}
&\multicolumn{1}{c}{$E_{val}$} & \multicolumn{1}{c}{$E_{tot}$}
&\multicolumn{1}{c}{BE/fermion} & \multicolumn{1}{c}{$S(0)$} \\
\multicolumn{8}{l}{ Kink Solution } \\
 3 & 2.0 & 2.6741 & 1.4699 & 0      & 4.1442 & +0.3814 &\        \\
\  & 1.0 & 1.8892 & 1.4160 & 0      & 3.3054 & +0.1018 &\        \\
\  & 0.5 & 1.3333 & 1.3977 & 0      & 2.8377 & -0.0896 &\        \\
\multicolumn{8}{c}{\ } \\
 6 & 2.0 & 2.6919 & 2.9152 & 0      & 5.6072 & -0.0655 &\        \\
\  & 1.0 & 1.8962 & 2.8222 & 0      & 4.7188 & -0.2315 &\        \\
\  & 0.5 & 1.3333 & 2.7954 & 0      & 4.1292 & -0.3118 &\        \\
\multicolumn{8}{c}{\ } \\
18 & 2.0 & 2.8026 & 8.5657 & 0      &11.3691 & -0.3684 &\        \\
\  & 1.0 & 1.9253 & 8.4173 & 0      &10.3438 & -0.4254 &\        \\
\  & 0.5 & 1.3333 & 8.3862 & 0      & 9.7210 & -0.4599 &\        \\
\multicolumn{8}{c}{\ } \\
\multicolumn{8}{l}{ Shallow Bag Solution } \\
 3 & 2.0 & 0.1636 & 0.0035 & 2.7570 & 2.9241 & -0.0253 & -0.1606 \\
\  & 1.0 & 0.5181 & 0.0758 & 2.1243 & 2.7182 & -0.0939 & -0.5146 \\
\  & 0.5 & 0.6586 & 0.3157 & 1.3990 & 2.3732 & -0.2089 & -0.8463 \\
\multicolumn{8}{c}{\ } \\
 6 & 2.0 & 1.4170 & 0.3683 & 3.5307 & 5.3159 & -0.1140 & -0.7810 \\
\  & 1.0 & 1.4045 & 1.0961 & 1.9963 & 4.4969 & -0.2505 & -1.1395 \\
\  & 0.5 & 0.9871 & 1.4204 & 1.5137 & 3.9212 & -0.3465 & -1.2051 \\
\multicolumn{8}{c}{\ } \\
18 & 2.0 & 2.7619 & 6.6004 & 1.9628 &11.3250 & -0.3708 & -1.6553 \\
\  & 1.0 & 1.7976 & 6.2340 & 2.2075 &10.2391 & -0.4312 & -1.5307 \\
\  & 0.5 & 1.1781 & 6.2803 & 2.1361 & 9.5945 & -0.4670 & -1.4884 \\
\end{tabular}
\label{tabd}
\end{table}

\newpage
\begin{table}
\caption{ Same as for Tables \ref{tabc}\ and \ref{tabd}\ but for 3-fermion
	``baryons''.  See text for discussion. }
\begin{tabular}{ccccccc}
\multicolumn{1}{c}{Solution} & \multicolumn{1}{c}{$\lambda/g^2$}
&\multicolumn{1}{c}{$E_{meson}$} &\multicolumn{1}{c}{$E_{loop}$}
&\multicolumn{1}{c}{$E_{val}$} & \multicolumn{1}{c}{$E_{tot}$}
&\multicolumn{1}{c}{BE/fermion} \\
\multicolumn{7}{l}{ Minimal Renormalization} \\
\hfil Shallow Bag& 2.0 & 0.0506 & 0.0249 & 2.8928 & 2.9683 & -0.0106 \\
\                & 1.0 & 0.0788 & 0.0779 & 2.7681 & 2.9247 & -0.0251 \\
\                & 0.5 & 0.0878 & 0.1660 & 2.6173 & 2.8711 & -0.0430 \\
\multicolumn{7}{c}{\ } \\
\hfil Kink       & 2.0 & 1.3333 & 1.9099 & 0      & 3.2433 & +0.0811 \\
\                & 1.0 & 0.9521 & 1.9230 & 0      & 2.8753 & -0.0416 \\
\                & 0.5 & 0.7079 & 1.9482 & 0      & 2.6562 & -0.1146 \\
\multicolumn{7}{c}{\ } \\
\multicolumn{7}{l}{ Full Renormalization } \\
\hfil Shallow Bag& 2.0 & 0.1473 & 0.0058 & 2.7742 & 2.9273 & -0.0242 \\
\                & 1.0 & 0.3371 & 0.0699 & 2.3615 & 2.7685 & -0.0772 \\
\                & 0.5 & 0.3633 & 0.2136 & 1.9746 & 2.5515 & -0.1495 \\
\multicolumn{7}{c}{\ } \\
\hfil Kink       & 2.0 & 1.3459 & 1.4576 & 0      & 2.8036 & -0.0655 \\
\                & 1.0 & 0.9481 & 1.4111 & 0      & 2.3594 & -0.2135 \\
\                & 0.5 & 0.6666 & 1.3977 & 0      & 2.0646 & -0.3118 \\
\end{tabular}
\label{tabe}
\end{table}


\begin{references}
\bibitem[1]{js}{ J.Schwinger, Ann. Phys. (N.Y.) {\bf 2}, 407 (1959).}
\bibitem[2]{gl}{ M.Gell-Mann and M.Levy, Nuovo Cimento {\bf 16}, 706 (1960).}
\bibitem[3]{lee}{ B.W.Lee, \underbar{Chiral Dynamics}, Gordon and Breach,
	N.Y. (1972).}
\bibitem[4]{candl}{ D.Campbell and Y.-T.Liao, Phys. Rev. {\bf D14},
	2093 (1976).}
\bibitem[5]{col}{ S.Coleman, ``Classical Lumps and Their Quantum Descendents,''
	in \underbar{Aspects of Symmetry},
	Cambridge University Press, Cambridge (1985).}
\bibitem[6]{sbag}{ W.A.Bardeen, M.S.Chanowitz, S.D.Drell, M.Weinstein and
	T.-N.Yan, Phys. Rev. {\bf D11}, 1094 (1975).}
\bibitem[7]{lw}{ T.D.Lee and G.C.Wick, Phys. Rev. {\bf D9}, 2291 (1974).}
\bibitem[8]{bog}{ See, {\it e.g.}, J.Boguta, Nucl. Phys. {\bf A501}, 637
(1989).}
\bibitem[9]{ahh}{ G.W.Anderson, L.J.Hall and S.D.H.Hsu, Phys.Lett.{\bf
	B249}, 505 (1990).}
\bibitem[10]{dlst}{ S.Dimopolous, B.W.Lynn, S.Selipsky and N.Tetradis,
	Phys. Lett. {\bf B253}, 237 (1991).}
\bibitem[11]{bandn}{ J.A.Bagger and S.G.Naculich, Phys. Rev. Let. {\bf 67},
2252
	(1991).}
\bibitem[12]{lpw}{ M.Li, R.J.Perry and L.Wilets, Phys. Rev. {\bf D36},
	596 (1987);
	see also: M.Li and R.J.Perry, Phys. Rev. {\bf D37}, 1670 (1988) ;
	M.Li, L.Wilets and R.J.Perry, Jour. Comp. Phys. {\bf 85}, 457 (1989). }
\bibitem[13]{wass}{ D.A.Wasson, Nucl. Phys. {\bf A535}, 456 (1991);
	see also: D.A.Wasson and S.Koonin, Phys. Rev. {\bf D43},3400
	(1991). }
\bibitem[14]{fer}{ T.C.Ferr\'ee, Ph.D. Thesis, University of Colorado,
	1992, unpublished; T.C.Ferr\'ee, C.E.Price and J.R.Shepard,
	Phys. Rev. {\bf C}, to be published.}
\end{references}
\end{document}